\documentclass[twocolumn,showpacs,aps,prd,superscriptaddress]{revtex4}

\usepackage{graphicx}
\usepackage{dcolumn}
\usepackage{amsmath}
\usepackage{epsfig}
\usepackage{bm}

\input babarsym

\newcommand{\BABARPubYear}    {06}
\newcommand{\BABARPubNumber}  {006}

\newcommand{\SLACPubNumber} {11839}

\newcommand{\kppcc}{\ensuremath{K^{\pm}\pi^{\pm}\pi^{\mp}}}
\newcommand{\btokppcc}{\ensuremath{\Bpm\to\kppcc}} 

\newcommand{\kkkcc}{\ensuremath{K^{\pm}K^{\pm}K^{\mp}}}
 
\newcommand{\btokkkcc}{\ensuremath{\Bpm\to\Kpm\Kpm\Kmp}}

\newcommand{\xz}{\ensuremath{X_0(1550)}}
\newcommand{\chiLLR}{\ensuremath{\chi^2_{\rm{LLR}}}}
\newcommand{\tc}[1]{\multicolumn{2}{c}{#1}}
\newcommand{\thc}[1]{\multicolumn{3}{c}{#1}}
\begin{document}

\begin{flushleft}
SLAC-PUB-\SLACPubNumber \\
\babar-PUB-\BABARPubYear/\BABARPubNumber
\end{flushleft}

\title{
{
\large \bf \boldmath Dalitz plot analysis of the decay $B^{\pm}\to K^{\pm}K^{\pm}K^{\mp}$
}
}

%
\author{B.~Aubert}
\author{R.~Barate}
\author{M.~Bona}
\author{D.~Boutigny}
\author{F.~Couderc}
\author{Y.~Karyotakis}
\author{J.~P.~Lees}
\author{V.~Poireau}
\author{V.~Tisserand}
\author{A.~Zghiche}
\affiliation{Laboratoire de Physique des Particules, F-74941 Annecy-le-Vieux, France }
% ...
\collaboration{The \babar\ Collaboration}
\noaffiliation

\date{\today}

\begin{abstract}
We analyze the three-body charmless decay $B^{\pm}\to K^{\pm}K^{\pm}K^{\mp}$
using a sample of $226.0 \pm 2.5$ million $B{\kern 0.18em\overline{\kern -0.18em B}{}\xspace}$ 
pairs collected by the 
{\mbox{\slshape B\kern-0.1em{\smaller A}\kern-0.1em B\kern-0.1em{\smaller A\kern-0.2em R}}} detector.
We measure the total branching fraction and $C\!P$ asymmetry to be 
${\cal B} = (35.2 \pm 0.9 \pm 1.6) \times 10^{-6}$ and $A_{C\!P} = (-1.7 \pm 2.6 \pm 1.5)\%$. 
We fit the Dalitz plot distribution using an isobar model and measure 
the magnitudes and phases of the decay coefficients.  
We find no evidence of $C\!P$ violation for the individual components of the isobar model.
The decay dynamics is dominated by the $K^{+}K^{-}$ $S$-wave, for which we perform a 
partial-wave analysis in the region $m(K^{+}K^{-}) < 2 {{\mathrm{\,Ge\kern -0.1em V\!/}c^2}}\xspace$. 
Significant production of the $f_0(980)$ resonance, and of a spin zero state near 
$1.55 {{\mathrm{\,Ge\kern -0.1em V\!/}c^2}}\xspace$ are required in the isobar model description of the data. 
The partial-wave analysis supports this observation.

\end{abstract}


\maketitle

\section{Introduction}

Charmless decays of $B$ mesons provide a rich laboratory for 
studying different aspects of weak and strong interactions. 
With recent theoretical progress in understanding the 
strong interaction effects,
specific predictions for two-body pseudoscalar-pseudoscalar, $B\to PP,$ 
and pseudoscalar-vector, $B\to PV$, branching fractions and 
asymmetries are available
~\cite{PRD-65-094025-Du,NPB-675-333-Beneke,PRD-64-112002-Chen,PLB-542-Colangelo} 
and global fits to experimental data have been performed
~\cite{PRD-67-014023-Du,PRD-68-113005-deGroot}.
Improved experimental measurements of a comprehensive set of charmless $B$ decays coupled
with further theoretical progress hold the potential to
provide significant constraints on the CKM matrix parameters and to discover hints
of physics beyond the Standard Model in penguin-mediated $b\to s$ transitions.

We analyze the decay $\btokkkcc$, dominated by
the $b\to s$ penguin-loop transition,
using $226.0 \pm 2.5$ million $\BB$ pairs 
collected by 
the \babar\ detector~\cite{babar} at the SLAC PEP-II asymmetric-energy 
$B$ factory~\cite{pep2cdr} 
operating at the $\FourS$ resonance. 
\babar\ has previously measured the total branching fraction and asymmetry in 
this mode~\cite{PRL-91-051801-Back} and the two-body branching fractions $\B^{\pm}\to K^{\pm}\phi(1020)$, 
$\B^{\pm}\to K^{\pm}\chi_{c0}$~\cite{PRD-69-01102-Telnov,PRD-69-071103-Tosi}. 
A comprehensive Dalitz plot analysis of $\btokkkcc$ has been 
published by the Belle collaboration~\cite{PRD-71-092003-Garmash}.

\section{Event selection}

We consider events with at least four 
reliably reconstructed charged-particle tracks 
consistent with having originated from the interaction point. 
All three tracks forming a \btokkkcc\ decay candidate are 
required to be consistent with a kaon hypothesis using 
a particle identification algorithm that has an average efficiency
of 94\% within the acceptance of the detector
and an average pion-to-kaon misidentification probability of 6\%.

We use two kinematic variables to identify the signal. 
The first is $\DeltaE = E - \sqrt{s_0}/2$,
the difference between the reconstructed $B$ candidate energy and 
half the energy of the $\epem$ initial state, both in the $\epem$ center-of-mass (CM) frame.
For signal events the $\DeltaE$ distribution peaks near zero 
with a resolution of $21\mev$. We require the candidates to have $|\DeltaE| < 40 \mev$.
The second variable is the energy-substituted mass 
$\mes = \sqrt{(s_0/2 + \bm{p}_0\cdot \bm{p}_B)^2/E^2_0 - \bm{p}^2_B}$,
where $\bm{p}_B$ is the momentum of the $B$ candidate and 
$(E_0,\bm{p}_0)$ is the four-momentum of the $\epem$ initial state, both in the laboratory frame.
For signal events the $\mes$ distribution peaks near the $B$ mass with 
a resolution of $2.6 \mevcc$. We define 
a signal region (SR) with $\mes\in(5.27,5.29) \gevcc$
and a sideband (SB) with $\mes\in(5.20,5.25) \gevcc$.

The dominant background is due to events from light-quark or charm 
continuum production, $\epem\to\qqbar$, whose jet-like event topology is different from the more spherical 
$B$ decays. We suppress this continuum background by requiring 
the absolute value of the cosine of the angle between the thrust axes of the $B$ candidate
and the rest of the event in the CM frame to be smaller than 0.95. 
Further suppression is achieved using a neural network with four inputs computed in the CM frame:
the cosine of the angle between the direction of the $B$ candidate and the beam direction; 
the absolute value of the cosine of the angle between the candidate thrust axis and the beam direction; 
and momentum-weighted sums over tracks and neutral clusters not belonging to the candidate,
$\sum_i p_i$ and $\sum_i p_i\cos^2\theta_i$, where $p_i$ is the track momentum and 
$\theta_i$ is the angle between the track momentum direction and the candidate thrust axis.

\begin{figure}[htb]
\begin{center}
\includegraphics[width=\columnwidth]{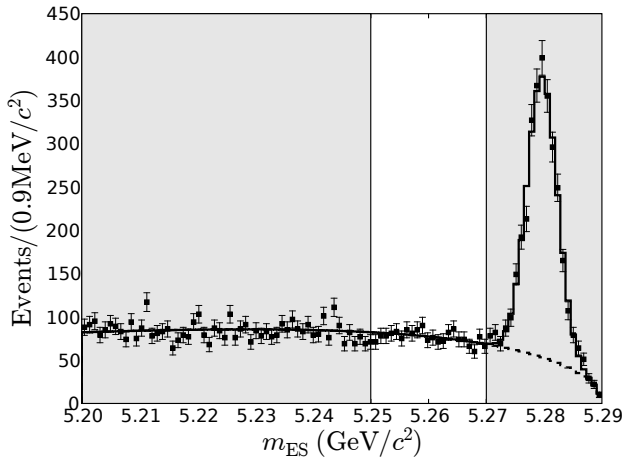}
\caption{The \mes distribution of the 9870 selected events, shown as the 
data points with statistical errors. The solid histogram shows 
a fit with a sum of a Gaussian distribution 
($m_0 = 5.2797 \pm 0.0007 \gevcc$, $\sigma = 2.64 \pm 0.07 \mevcc$, $N = 2394 \pm 63$) 
and a background function~\cite{ZPC-48-543-ARGUS}, shown as a dashed histogram. The 
shaded regions correspond to the signal region and the $\mes$ sideband defined in the text.}
\label{mesfit}
\end{center}
\end{figure}

Figure~\ref{mesfit} shows the $\mes$ distribution of the 9870 events thus selected.
The histogram is fitted with a sum of a Gaussian distribution and a background function 
having a probability density,
$P(x) \propto x\sqrt{1-x^2}\exp{(-\xi(1-x^2))}$,
where $x=2\mes/\sqrt{s_0}$ and $\xi$ is a shape parameter~\cite{ZPC-48-543-ARGUS}. 
The binned maximum likelihood fit gives $\chi^2 = 104$ for 100 bins and $\xi = 21.1 \pm 1.6$.
The ratio of the integrals of the background function over the signal region 
and the sideband yields
an extrapolation coefficient $R_{\qqbar} = 0.231 \pm 0.007$. The 
expected number of $\qqbar$ background events in the signal region is 
$n^{\rm SR}_{\qqbar} = R_{\qqbar}(n^{\rm SB} - n^{\rm SB}_{\BB}) = 972 \pm 34$, 
where $n^{\rm SB} = 4659$ is the 
number of events in the sideband from which we subtract the number of 
non-signal $\BB$ background 
events $n^{\rm SB}_{\BB} = 431 \pm 19$, estimated  
using 
a large number of simulated exclusive $B$ decays. The expected number of $\BB$ background
events in the signal region is $n^{\rm SR}_{\BB} = 276 \pm 20$, with $\Bpm\to D\Kpm$ decays 
giving the largest contribution.

\begin{figure}[htb] 
\begin{center}
\includegraphics[width=\columnwidth]{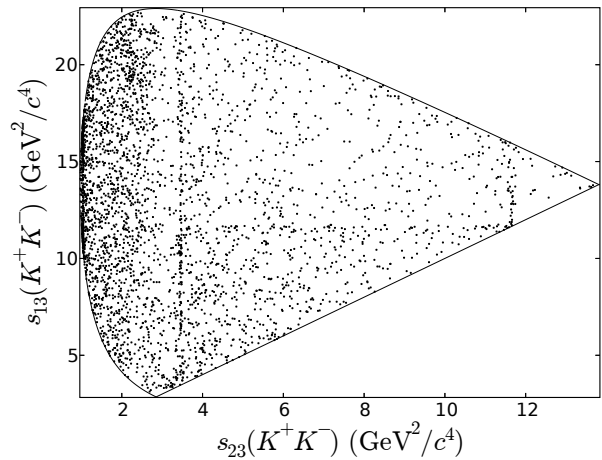}
\caption{The Dalitz plot of the 1769 $\Bp$ and 1730 $\Bm$ candidates selected in the 
signal $\mes$ region. The axes are defined in the text.}
\label{dalitzplot}
\end{center} 
\end{figure}

We use a kinematic fit, constraining the mass of the selected candidates to 
the mass of the $B$ meson. 
The three-body decay kinematics is 
described by two independent di-kaon invariant mass variables
$(m^2_{23},m^2_{13}) = (s_{23},s_{13})$,
where we order the same-sign kaons such that $s_{23} \le s_{13}$.
The signal region contains 1769 $\Bp$ and 1730 $\Bm$ candidates
whose Dalitz plot distribution is shown in Fig.~\ref{dalitzplot}. 

\section{Isobar model fit}

We perform an extended binned maximum likelihood fit to the 
event distribution in Fig.~\ref{dalitzplot} by binning
the folded Dalitz plot into 292 non-uniform rectangular bins and 
minimizing the log of the Poisson likelihood ratio, 
$\chiLLR/2 = \sum^{292}_{i=1} \mu_i - n^{\rm SR}_i + n^{\rm SR}_i\ln\left({n^{\rm SR}_i/\mu_i}\right)$, 
where $n^{\rm SR}_i$ is the number of observed signal region events in the $i$-th bin, 
assumed to be sampled from a Poisson distribution with mean $\mu_i$. 
In the limit
of large statistics, the $\chiLLR$ function has a $\chi^2$ distribution and can be used to evaluate
the goodness-of-fit. 

The expected number of events in the $i$-th bin is modeled as
\begin{equation}
\mu_i = 2\int_i \epsilon |{\cal M}|^2 dm^2_{23}dm^2_{13} + R_{\qqbar}(n^{\rm SB}_i - n^{\rm SB}_{i\BB}) + n^{\rm SR}_{i\BB},
\end{equation}
where the first term is the expected signal contribution, given by twice the bin integral 
of the square of the matrix element multiplied by the signal selection 
efficiency $\epsilon$, 
determined as a function of $(m_{23},m_{13})$ using simulated signal events. The integral
is multiplied by two because we use a folded Dalitz plot.   
We use the isobar model formalism~\cite{PR-135-B551-Fleming,PR-166-1731-Morgan}
and describe the matrix element $\cal M$ as a sum of 
coherent contributions, ${\cal M} = \sum^{N}_{k = 1} {\cal M}_{k}$.
The individual contributions are symmetrized with respect to 
the interchange of same-sign kaons, ${1\leftrightarrow 2}$, and are given by
\begin{equation}
{\cal M}_k = \frac{\rho_k e^{i\phi_k}}{\sqrt{2}}
\left({\cal A}_k(s_{23})P_{J_k}(\cos\theta_{13}) + \{1 \leftrightarrow 2\}\right),
\end{equation}
where $\rho_k e^{i\phi_k}$ is a complex-valued decay coefficient,
${\cal A}_k$ is the amplitude describing a $\Kp\Km$ system in a state
with angular momentum $J$ and invariant mass $\sqrt{s_{23}}$,
$P_{J}$ is the Legendre polynomial of order $J$, and the helicity angle $\theta_{13}$ 
between the direction of the bachelor recoil kaon 1 and kaon 3 
is measured in the rest frame of kaons 2 and 3.

The model includes contributions from the $\phi(1020)$ and $\chi_{c0}$ 
intermediate resonances, which are clearly visible in Fig.~\ref{dalitzplot}. 
Following Ref.~\cite{PRD-71-092003-Garmash}, we introduce a broad scalar resonance, 
whose interference with a slowly varying nonresonant component 
is used to describe the rapid decrease in event density around $m(\Kp\Km) = 1.6\gevcc$. 
Evidence of a possible resonant $S$-wave contribution in this region has been reported previously
\cite{ZPC-17-309-Baubillier,NPB-B301-525-Aston}, however its attribution is uncertain:
the $f_0(1370)$ and $f_0(1500)$ resonances are 
known to couple more strongly to $\pi\pi$ than to $K\Kbar$ \cite{PDG};
possible interpretations in terms of those states~\cite{EPJ-C39-71-Minkowski}
must account for the fact that 
no strong $B^{\pm}\to K^{\pm} f_0(1370)$ or $B^{\pm}\to K^{\pm} f_0(1500)$
signal is observed in $\btokppcc$~\cite{PRD-72-072003,PRD-71-092003-Garmash}.
The contribution of the $f_0(1710)$ resonance
is included in the fit as a separate component and is found to be small. In the following, 
we designate the broad scalar resonance $\xz$ and determine its mass and width directly from the fit. 

The contribution from a spin $J$ resonance with mass $m_0$ and 
total width $\Gamma_0$ is described by a relativistic Breit-Wigner amplitude:
\begin{equation}
{\cal A}_J(s) = \frac{F_J(q_{\Kpm}R)}{m^2_0 - s - im_0(\Gamma_0 + \Delta\Gamma(s))}. 
\label{breitwigner}
\end{equation}
$F_J$ is the Blatt-Weisskopf centrifugal barrier factor~\cite{blattweisskopf} for
angular momentum $J$: $F_0(x) \equiv 1$ and $F_1(x) \equiv x/\sqrt{1+x^2}$, 
$q_{h} = \sqrt{s/4-m^2_{h}}$, and 
$R$ represents the effective radius of the interaction volume for the resonance; 
we use $R = 4.0 \gev^{-1}$ ($0.8 \fm$)~\cite{PLB-621-72-2005}. 
In the formulation of Eq.~(\ref{breitwigner}), only the centrifugal barrier factor
for the decay of a spin $J$ resonance into two pseudoscalar kaons is included; we have ignored the corresponding centrifugal 
barrier factor for the two-body decay of a $B$ meson into a pseudoscalar kaon
and a spin $J$ resonance. The effect of this approximation on the parameterization
of $\Bpm\to\Kpm\phi(1020)$, the only component with $J > 0$, is negligible. 
Unless otherwise specified, all resonance parameters are taken from Ref.~\cite{PDG}.
The term $\Delta\Gamma(s)$, parameterizing the mass dependence of the total width,
is in general given by $\Delta\Gamma(s) = \sum_i \Delta\Gamma_i(s)$, 
where the sum is over all decay modes of the resonance, and $\Delta\Gamma_i(m^2_0) \equiv 0$. 
The $\chi_{c0}$  has many decay modes, the decay modes of the $f_0(1710)$ are not
well established, and decay modes other than  $\KpKm$ of the possible $\xz$ resonance are unknown;
in all these cases we set $\Delta\Gamma(s) = 0$ and neglect the mass dependence of the total width. For the 
$\phi(1020)$ resonance we use $\Delta\Gamma(s) = \Delta\Gamma_1(s) + \Delta\Gamma_2(s)$,
where $\Gamma_1 = \Gamma_0\BR(\phi\to\Kp\Km)$, $\Gamma_2 = \Gamma_0\BR(\phi\to\Kz\Kzb)$, and
the mass dependence of the partial width for the two-body vector to pseudoscalar-pseudoscalar decay
$\phi(1020) \to h\bar{h}$ is parameterized as
\begin{equation}
\Delta\Gamma_{1,2}(s) = 
\Gamma_{1,2}\left(\frac{q_h}{q_{0h}}\frac{m_{\phi}}{\sqrt{s}}\frac{F^2_1(q_hR)}{F^2_1(q_{0h}R)} - 1\right), 
\end{equation}
where $q_{0h} = \sqrt{m^2_\phi/4-m^2_{h}}$.

\begin{figure}[htb]
\begin{center}
\includegraphics[width=\columnwidth]{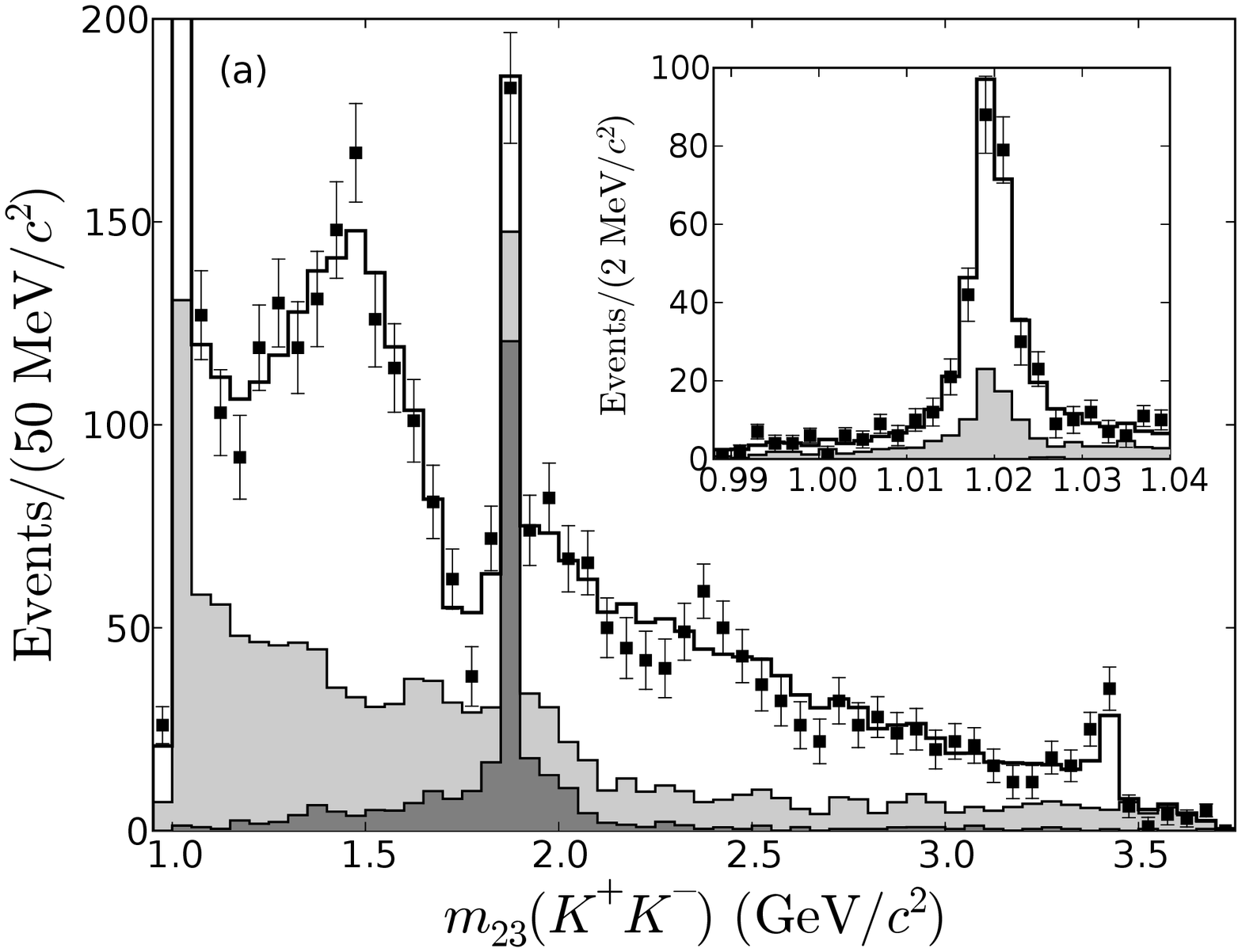} \\
\includegraphics[width=\columnwidth]{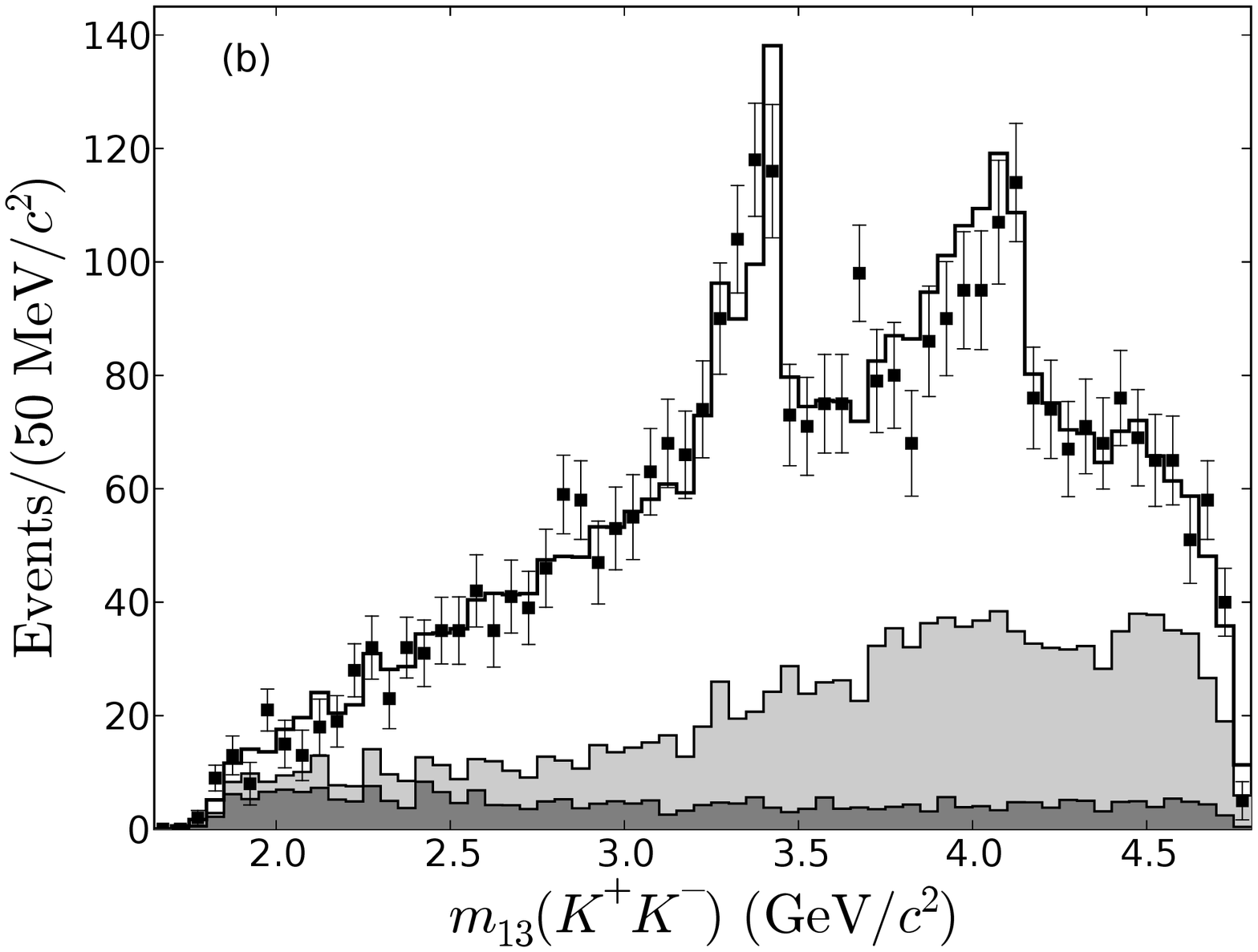} 
\caption{The projected $m(\Kp\Km)$ invariant-mass distributions for the best fit: 
(a) $m_{23}$ projection (the inset shows the fit projection near the $\phi(1020)$ resonance),
(b) $m_{13}$ projection.
The histograms show the result of the fit with $\BB$ and $\qqbar$ background contributions shown in dark and
light gray, respectively.}
\label{fitproj}
\end{center}
\end{figure}

A large $B^{\pm}\to K^{\pm} f_0(980)$
signal measured in $\btokppcc$~\cite{PRD-72-072003,PRD-71-092003-Garmash},
and a recent measurement of $g_{K}/g_{\pi}$, 
the ratio of the $f_0(980)$ coupling 
constants to $K\Kbar$ and 
$\pi\pi$~\cite{PLB-607-243-BES-2005}, 
motivate us to include  
an $f_0(980)$ contribution using a coupled-channel amplitude parameterization:
\begin{equation}
{\cal A}_{f_0(980)}(s) = \frac{1}
{m^{2}_{0} - s - im_{0}
\left(g_{\pi}\varrho_{\pi} + g_{K}\varrho_{K}\right)},
\label{flatte}
\end{equation}
where $\varrho_{\pi}=2/3 \sqrt{1-4m_{\pipm}^2/s} + 1/3\sqrt{1-4m_{\piz}^2/s}$,
$\varrho_{K}=1/2 \sqrt{1-4m_{\Kpm}^2/s} + 1/2\sqrt{1-4m_{\Kz}^2/s}$,
and we use $g_{K}/g_{\pi} = 4.21 \pm 0.25 \pm 0.21$, $m_0 = 0.965 \pm 0.008 \pm 0.006 \gevcc$ and
$g_{\pi} = 0.165 \pm 0.010 \pm 0.015 \gevcc$~\cite{PLB-607-243-BES-2005}.

We have investigated two theoretical models
of the nonresonant component~\cite{PRD-66-054015-Cheng,PRD-70-034033-Fajfer}  
and found that neither describes the data adequately.
We therefore include an $S$-wave nonresonant component
expanded beyond the usual constant term as
\begin{equation}
{\cal M}_{\rm NR} = \frac{\rho_{\rm NR} e^{i\phi_{\rm NR}}}{\sqrt{2}} 
\left(e^{-(\alpha + i\beta) s_{23}} + e^{-(\alpha + i\beta) s_{13}} \right).
\label{nrexpansion}
\end{equation}
A fit to the $m_{23} > 2\gevcc$ region of the folded Dalitz plot of Fig~\ref{dalitzplot}, 
which is dominated by the nonresonant component,
gives $\alpha = 0.140 \pm 0.019 \gev^{-2}c^4$, $\beta = -0.02 \pm 0.06 \gev^{-2}c^4$, 
consistent with no phase variation. In the following we fix $\beta = 0$, and incorporate
the ${\cal M}_{\rm NR}$ contribution over the entire Dalitz plot,
thus effectively employing 
the same parameterization as in Ref.~\cite{PRD-71-092003-Garmash}.

We fit for the magnitudes and phases of the decay coefficients, the
mass and width of the $\xz$, and the nonresonant component shape parameter $\alpha$. 
As the overall complex phase of the isobar model amplitude is arbitrary,
we fix the phase of the nonresonant contribution to zero, leaving 14 free parameters in the fit.
The number of degrees of freedom is $292 - 14 = 278$.  
We perform multiple minimizations with different starting points and find multiple 
solutions clustered in pairs, where the solutions within each pair 
are very similar, except for
the magnitude and phase of the $\chi_{c0}$ decay coefficient. The 
twofold ambiguity arises from the interference between the narrow $\chi_{c0}$ 
and the nonresonant component, which is approximately constant across the resonance. 
The highest-likelihood pair has $\chiLLR = (346.4,352.0)$; the second best
pair has $\chiLLR = (362.4,368.7)$. The least significant components 
are the $f_0(980)$ and the $f_0(1710)$. Their omission from the fit model degrades the 
best fit from $\chiLLR = 346.4$ to $363.9$ and $360.7$, respectively.
\begin{table*}[htb]
\caption{The magnitudes and phases of the decay coefficients, fit fractions, two-body
branching fractions, $\CP$ asymmetries, symmetric 90\% confidence level $\CP$ asymmetry intervals
around the nominal value, and the phase differences between the charge-dependent decay coefficients 
for the individual components of the isobar model fit.}
\vspace{1mm}
\resizebox{\textwidth}{!}{
\begin{tabular}{c 
r@{$\ \pm \ $}l  
r@{$\ \pm \ $}c@{$\ \pm \ $}l 
r@{$\ \pm \ $}c@{$\ \pm \ $}l
r@{$\ \pm \ $}c@{$\ \pm \ $}l
r@{$\ \pm \ $}c@{$\ \pm \ $}l
c
r@{$\ \pm \ $}c@{$\ \pm \ $}l
}
\hline
\hline
                    Comp. & 
              \tc{$\rho$} &  
             \thc{$\phi$ (rad)} & 
                \thc{$F$ (\%)} & 
                \thc{$F\times\BR(\btokkkcc)$} & 
                \thc{$A$} & 
                 $(A_{\rm min},A_{\rm max})_{90\%}$ &
\thc{$\delta\phi$ (rad)} \\
\hline
                 $\phi(1020)$  &  
              $1.66$ & $0.06$ &
              $2.99$ & $0.20$ & $0.06$ &  
     $11.8$ & $0.9$ & 0.8  & 
(4.14 & 0.32 & 0.33)$\times 10^{-6}$ &
     $0.00$ & $0.08$ & 0.02  &
         $(-0.14,0.14)$  &
     $-0.67$ & $0.28$ & 0.05  \\
             $f_0(980)$  &  
               $5.2$ & $1.0$ &
    $0.48$ & $0.16$ & $0.08$  &  
    $19$ & $7$ & 4  & 
(6.5 & 2.5 & 1.6)$\times 10^{-6}$ &
    $-0.31$ & $0.25$ & 0.08  &
         $(-0.72,0.12)$  &
    $-0.20$ & $0.16$ & 0.04  \\
                  $\xz$  &  
            $8.2$ & $1.1$ & 
    $1.29$ & $0.10$ & $0.04$  & 
          $121$ & $19$ & 6  &
(4.3 & 0.6 & 0.3)$\times 10^{-5}$ & 
    $-0.04$ & $0.07$ & 0.02  &
         $(-0.17,0.09)$  &
    $0.02$ & $0.15$ & 0.05  \\
            $f_0(1710)$  &   
           $1.22$ & $0.34$ & 
    $-0.59$ & $0.25$ & $0.11$  &   
      $4.8$ &  $2.7$ & 0.8  & 
(1.7 & 1.0 & 0.3)$\times 10^{-6}$ &
    $0.0$ & $0.5$ & 0.1  &
         $(-0.66,0.74)$  &
    $-0.07$ & $0.38$ & 0.08  \\
            $\chi^{I}_{c0}$  &   
            $0.437$ & $0.039$ &
     $-1.02$ & $0.23$ & $0.10$  &   
       $3.1$ & $0.6$ & $0.2$  & 
(1.10 & 0.20 & 0.09)$\times 10^{-6}$ &
     $0.19$ & $0.18$ & 0.05  &
         $(-0.09,0.47)$  &
     $0.7$ & $0.5$ & 0.2  \\
            $\chi^{II}_{c0}$  &   
           $0.604$ & $0.034$ &
\thc{$0.29 \pm 0.20$} &
\thc{$6.0 \pm 0.7$} & 
\thc{$(2.10 \pm 0.24) \times 10^{-6}$} &
\thc{$-0.03 \pm 0.28$} &
- &
\thc{$-0.4 \pm 1.3$} \\
                     NR  &   
             $13.2$ & $1.4$ & 
   \thc{0}  & 
          $141$ & $16$ & 9  & 
(5.0 & 0.6 & 0.4)$\times 10^{-5}$ &
     $0.02$ & $0.08$ & 0.04  &
         $(-0.14,0.18)$  & 
   \thc{0}  \\
\hline
\hline
\end{tabular}
}
\label{fitresults}
\end{table*}

The invariant-mass projections of the best fit are shown in Fig.~\ref{fitproj}. The goodness-of-fit is 
$\chi^2 = 56$ for 56 bins in the $m_{23}$ projection and $\chi^2 = 66$ for 63 bins in the $m_{13}$ 
projection.
The sharp peak in the $\BB$ background 
distribution in the $m_{23}$ projection is due to the contribution from the $B^{\pm}\to D\Kpm$
backgrounds.
The fit gives $\alpha = 0.152 \pm 0.011 \gev^{-2}c^4$, $m_0(X_0) = 1.539 \pm 0.020 \gevcc$, and
$\Gamma_0(X_0) = 0.257 \pm 0.033 \gevcc$. 
The fitted values of the shape parameter $\alpha$ 
and the resonance mass are consistent with the values in Ref.~\cite{PRD-71-092003-Garmash},
but our preferred value for the width is significantly larger.  
The results of the best isobar model fit are summarized in Table~\ref{fitresults},
where we have also included the results for the $\chi_{c0}$ component 
from the second solution in the highest-likelihood pair, labeled $\chi^{II}_{c0}$, and
component fit fractions,
\begin{equation}
F_k = 
\frac{\int |{\cal M}_k|^2 ds_{23}ds_{13}}
{\int |{\cal M}|^2 ds_{23}ds_{13}},
\label{fitfractions}
\end{equation}
where the integrals are taken over the entire Dalitz plot. The sum 
of the component fit fractions is 
significantly larger than one due to 
large negative interference in the scalar sector~\cite{epaps}.

\section{Branching fractions and asymmetries}

To search for possible direct $\CP$ violation we extend the isobar model 
formalism by defining charge-dependent decay coefficients:
\begin{equation}
\rho^{\mp}_k e^{i\phi^\mp_k} = 
\rho_k e^{i\phi_k} \sqrt{\frac{1 \pm A_k}{2}} e^{\pm i\delta\phi_k/2},
\end{equation}
where $A_k$ is the $\CP$ asymmetry of the $k$-th component, and 
$\delta\phi_k = \phi^-_k - \phi^+_k$. 
We modify the likelihood to be the product of the likelihoods 
for the two charges and repeat the fit. The phase of the nonresonant
component is fixed to zero for both charges.
The results are given in 
the last three columns of Table~\ref{fitresults} in terms
of the fitted $\CP$ asymmetry values, 
the symmetric 90\% confidence level intervals around them, and
the fitted phase differences 
between the charge-dependent decay coefficients.
The asymmetry intervals are estimated by fitting Monte Carlo 
simulated samples generated
according to the parameterized model of the nominal asymmetry fit. There is no evidence
of statistically significant $\CP$ violation for any of the components.

Taking into account the signal Dalitz plot distribution, as described by the isobar model fit,
the average signal efficiency is $\bar{\varepsilon} = 0.282 \pm 0.011$, where
the uncertainty is evaluated using control data samples, and is 
primarily due to the uncertainties in tracking and 
particle identification efficiencies. 
The total branching fraction is
$\BR(\btokkkcc) =  (35.2 \pm 0.9 \pm 1.6) \times 10^{-6}$, where the first error is statistical
and the second is systematic.  
The fit fraction of the isobar model terms that do not involve the $\chi_{c0}$ resonance
is $(95.0 \pm 0.6 \pm 1.1)\%$ for the best fit, giving 
$\BR(\btokkkcc) =  (33.5 \pm 0.9 \pm 1.6) \times 10^{-6}$ if intrinsic charm 
contributions are excluded.
The total asymmetry is 
$A_{\CP} = \frac{\BR(\Bm\to\Km\Km\Kp) - \BR(\Bp\to\Kp\Kp\Km)}
{\BR(\Bm\to\Km\Km\Kp) + \BR(\Bp\to\Kp\Kp\Km)} = 
(-1.7 \pm 2.6 \pm 1.5)\%$.

The systematic error for the overall branching fraction is obtained by combining in quadrature
the $3.9\%$ efficiency uncertainty, a $1.1\%$ uncertainty on the total number of $\Bp\Bm$ pairs, 
a $0.7\%$ uncertainty due to the modeling of $\BB$ backgrounds, and a $1.4\%$ 
uncertainty due to the uncertainty arising from the uncertainty on the 
$R_{\qqbar}$ sideband extrapolation coefficient. 
The $1.5\%$ systematic uncertainty for the asymmetry 
is due to possible charge asymmetry in kaon tracking and 
particle identification efficiencies, evaluated using data control samples. 
Where appropriate, the systematic uncertainties discussed above have been propagated 
to estimate the uncertainties on the leading isobar model fit results. 
We have also evaluated the systematic uncertainties due to the parameterization 
of resonance lineshapes by varying the parameters of all resonances within their
respective uncertainties. Uncertainties arising from 
the distortion of narrow resonance lineshapes due to finite detector resolution and,
for candidates containing a $\phi(1020)$ resonance produced in the $\qqbar$ continuum,
due to the kinematic fit, have also been studied.

The values of the partial two-body branching fractions are summarized 
in the fifth column of Table~\ref{fitresults}. 
Using the $\BR(\phi(1020)\to\Kp\Km)$ and $\BR(\chi_{c0}\to\Kp\Km)$ branching fractions from Ref.~\cite{PDG}, 
we compute $\BR(\Bpm\to\Kpm\phi(1020)) = (8.4 \pm 0.7 \pm 0.7 \pm 0.1)\times 10^{-6}$ and 
$\BR(\Bpm\to\Kpm\chi_{c0}) = (1.84 \pm 0.32 \pm 0.14 \pm 0.28)\times 10^{-4}$, where the last
error is due to the uncertainty on the $\phi(1020)$ and $\chi_{c0}$ branching fractions. 
Both results are in agreement with previous 
measurements~\cite{PRL-86-3718,PRD-69-01102-Telnov,PRD-69-071103-Tosi,PRD-71-092003-Garmash,PRL-95-031801-Acosta}.

The partial branching fractions for $B^{\pm}\to K^{\pm} f_0(980)$ measured in 
the $\kkkcc$ and $\kppcc$ final states are related by the ratio
\begin{equation}
R = \frac{\BR(f_0(980)\to \Kp\Km)}{\BR(f_0(980)\to \pip\pim)} = 
\frac{3}{4}\frac{I_{K}}{I_{\pi}} \frac{g_{K}}{g_{\pi}}, 
\label{f0ratio}
\end{equation}
where 3/4 is an isospin factor, and $I_{K}/I_{\pi}$ is
the ratio of the integrals of the square of the $f_0(980)$ amplitude given by Eq.~(\ref{flatte})
over the $B\to KKK$ and $B\to K\pi\pi$ Dalitz plots, and $g_{K}/g_{\pi}$ is 
the ratio of the $f_0(980)$ coupling constants to $K\Kbar$ and $\pi\pi$. 
Using our results and those in Ref.~\cite{PRD-72-072003}, we get 
$R = 0.69 \pm 0.32$, where we have combined the statistical and systematic 
errors of the two measurements in quadrature. This is consistent with $R = 0.92 \pm 0.07$,
which we get by evaluating  
the right-hand side of Eq.~(\ref{f0ratio}) using the values of the $f_0(980)$ 
parameters reported by the BES collaboration~\cite{PLB-607-243-BES-2005}.

\section{Partial-wave analyses}

We further study the nature of the dominant $S$-wave component
by considering the interference between the low-mass 
and the high-mass scattering amplitudes in the region 
$m_{23}\in (1.1,1.8) \gevcc$, $m_{13} > 2 \gevcc$.
The matrix element is modeled as
\begin{equation}
{\cal M} = \frac{\rho_S(s_{23})}{\sqrt{2}}e^{i\phi_S(s_{23})} + \frac{\rho_{\rm NR}}{\sqrt{2}}e^{-\alpha s_{13}},
\end{equation}
where $\rho_S$ and $\phi_S$ refer to the $S$-wave and are taken to be constant within each bin
of the $s_{23}$ variable and the nonresonant amplitude parameterization is taken from the 
fit to the high-mass region.
The partial-wave expansion truncated at the $S$-wave describes the data adequately; 
the magnitude of the $S$-wave in each bin is readily determined.
Because of the mass dependence of the nonresonant component, 
the phase of the $S$-wave can also be determined, albeit with a sign ambiguity
and rather large errors for bins with a small number of entries or 
small net variation of the nonresonant component.

\begin{figure}[htb]
\begin{center}
\includegraphics[width=\columnwidth]{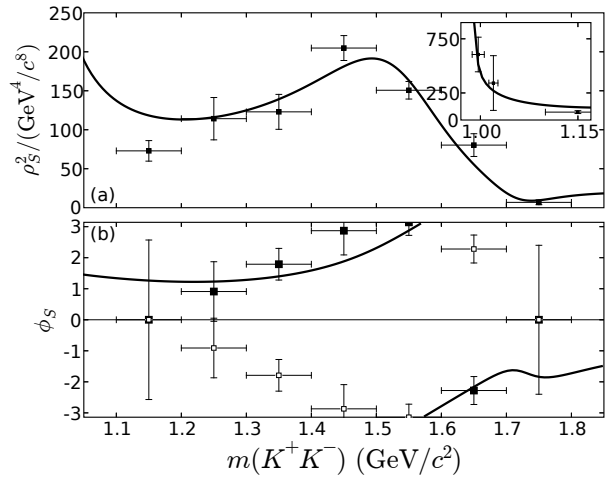}
\caption{The results of the partial-wave analysis of the $\Kp\Km$ $S$-wave: 
(a) magnitude squared, (b) phase. 
The discrete ambiguities in the determination of the 
phase give rise to two possible solutions labeled by black 
and white squares. The curves correspond to the $S$-wave component from the isobar model fit.
The inset shows the evidence of a threshold enhancement from the fits of the $S$-wave in the 
vicinity of the $\Kp\Km$ threshold and in the region around the $\phi(1020)$ resonance.
}
\label{swave}
\end{center}
\end{figure}
The results are shown in Fig.~\ref{swave}, with the $S$-wave component of the isobar model
fit overlaid for comparison. 
Continuity requirements allow us to identify two possible solutions
for the phase; the solution labeled by black squares is
consistent with a rapid counterclockwise motion in the Argand plot around $m(\Kp\Km) = 1.55\gevcc$, 
which is accommodated in the isobar model as the contribution of the $\xz$.

\begin{figure}[htb]
\begin{center}
\includegraphics[width=\columnwidth]{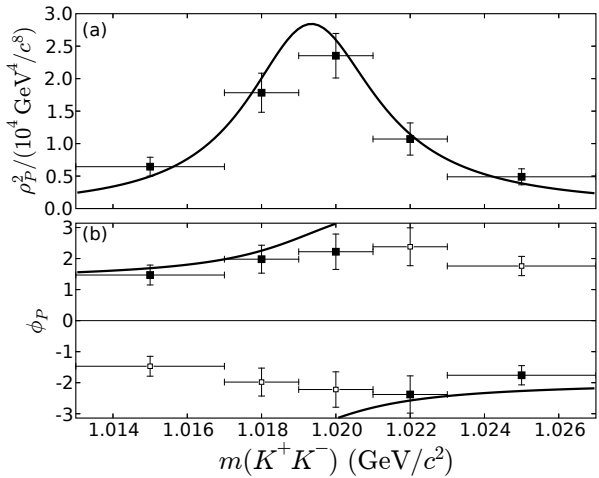}
\caption{
The results of the partial-wave analysis in the $\phi(1020)$ region for the $P$-wave: 
(a) magnitude squared, (b) phase.
The discrete ambiguities in the determination of the 
phase give rise to two possible solutions labeled by black 
and white squares.
The curve corresponds to a Breit-Wigner fit of the $\phi(1020)$ resonance.}
\label{pwave}
\end{center}
\end{figure}

Isospin symmetry relates the measurements in $\btokkkcc$ and $\Bz\to\Kp\Km\KS$~\cite{PRD-72-094031-Gronau}. 
Our results for the $\Kp\Km$ $S$-wave can therefore be used to estimate
a potentially significant source
of uncertainty in the measurements of $\stwob$ in 
$\Bz\to \phi(1020)\KS$~\cite{PRD-71-091102,PRL-91-261602}
due to the contribution of a $\CP$-even $S$-wave amplitude.
We perform a partial-wave analysis in the region $m_{23}(\Kp\Km) \in (1.013,1.027) \gevcc$, which
we assume to be dominated by the low-mass $P$-wave, due to the contribution of the $\phi(1020)$
resonance, and a low-mass $S$-wave. The matrix element is modeled as 
\begin{equation}
{\cal M} = \frac{\rho_S}{\sqrt{2}} + \frac{\rho_P(s_{23})}{\sqrt{2}}e^{i\phi_P(s_{23})}\cos\theta_{13},
\end{equation}
where the low-mass $S$-wave is taken to be constant over the small $s_{23}$ interval 
considered. The fit results for the $P$-wave are shown in Fig.~\ref{pwave}, with a Breit-Wigner
fit of the $\phi(1020)$ resonance overlaid for comparison. For the $S$-wave
we get $\rho^2_S = (3.4 \pm 2.5)\times 10^2 \gev^{-4}c^8$ 
and compute its fraction in this region using Eq.~(\ref{fitfractions}) to be $(9 \pm 6)\%$.

We also consider the region $2m_{\Kp} < m(\Kp\Km) < 1.006 \gevcc$, in the immediate vicinity of the 
 $\Kp\Km$ threshold. The contribution of the $\phi(1020)$ resonance tail in this region is suppressed by 
the centrifugal barrier and is estimated to be smaller than 10\%. We fit $\rho^2_S = (6.1 \pm 1.6)\times 10^2 \gev^{-4}c^8$
for the magnitude of the $S$-wave in this region. The fits in the 
vicinity of the $\Kp\Km$ threshold and in the region around the $\phi(1020)$ resonance
indicate a threshold enhancement of the $S$-wave, 
which is accommodated in the isobar model by the contribution of the $f_0(980)$ resonance
as shown in the inset of Fig.~\ref{swave}.

\section{Conclusions}

In conclusion, we have measured the total branching fraction and 
the $\CP$ asymmetry in $\btokkkcc$.
An isobar model Dalitz plot fit and a partial-wave analysis of the $\Kp\Km$ $S$-wave 
show evidence of large contributions from a broad $\xz$ scalar resonance, 
a mass-dependent nonresonant component, and an $f_0(980)$ resonance. 
The ratio of $B^{\pm}\to K^{\pm} f_0(980)$ two-body branching fractions measured
by $\babar$ in $\btokppcc$ and $\btokkkcc$ is    
consistent with the measurement of $g_{K}/g_{\pi}$
by the BES collaboration, albeit with large errors.
Our isobar model fit results are substantially different from those obtained in Ref.~\cite{PRD-71-092003-Garmash}
due to the larger fitted width of the $\xz$ and the inclusion of the $f_0(980)$ component in the isobar model. 
Our results for the $\BR(\Bpm\to\Kpm\phi(1020))$ and $\BR(\Bpm\to\Kpm\chi_{c0})$ branching fractions
are in agreement with the previous results from $\babar$~\cite{PRD-69-01102-Telnov,PRD-69-071103-Tosi}, which
they supersede, and from other experiments~\cite{PRL-86-3718,PRD-71-092003-Garmash,PRL-95-031801-Acosta}. 
We have measured the $\CP$ asymmetries and the phase differences between the charge-dependent
decay coefficients for the individual components of the isobar model and found no evidence of direct $\CP$ violation.

We wish to dedicate this paper to Prof. Richard E. (Dick) Dalitz,
inventor of the ``Phase Space Plot'' (as he called it) with which
his name is so intimately linked, in tribute as much to his qualities
as a gentleman as to his gifts as a physicist.

We are grateful for the excellent luminosity and machine conditions
provided by our \pep2\ colleagues, 
and for the substantial dedicated effort from
the computing organizations that support \babar.
The collaborating institutions wish to thank 
SLAC for its support and kind hospitality. 
This work is supported by
DOE
and NSF (USA),
NSERC (Canada),
IHEP (China),
CEA and
CNRS-IN2P3
(France),
BMBF and DFG
(Germany),
INFN (Italy),
FOM (The Netherlands),
NFR (Norway),
MIST (Russia), and
PPARC (United Kingdom). 
Individuals have received support from CONACyT (Mexico), 
Marie Curie EIF (European Union),
the A.~P.~Sloan Foundation, 
the Research Corporation,
and the Alexander von Humboldt Foundation.


\end{document}